\documentclass[aps,pre,eqsecnum,twocolumn,showpacs]{revtex4}
\usepackage[dvips]{graphicx}
\usepackage{amsmath,bm}
\newcommand{\pp}[2]{\frac{\partial #1}{\partial #2}}
\renewcommand{\Re}{\mathop{\rm Re}}

\newcommand{\Tr}{\mathop{\rm Tr}}

\begin{document}
\title{Anomalous shell effect in the transition from a circular to
a triangular billiard}

\author{Ken-ichiro Arita$^{1,2}$ and Matthias Brack$^2$}
\affiliation{$^1$Department of Physics, Nagoya Institute of Technology,
             466-8555 Nagoya, Japan\\
$^2$Institute for Theoretical Physics, University of
             Regensburg, D-93040 Regensburg, Germany}

\received{November 5, 2007}
\revised{\today}

\begin{abstract}
We apply periodic orbit theory to a two-dimensional non-integrable
billiard system whose boundary is varied smoothly from a circular to
an equilateral triangular shape. Although the classical dynamics
becomes chaotic with increasing triangular deformation, it exhibits an
astonishingly pronounced shell effect on its way through the shape
transition. A semiclassical analysis reveals that this shell effect
emerges from a codimension-two bifurcation of the triangular periodic
orbit. Gutzwiller's semiclassical trace formula, using a global
uniform approximation for the bifurcation of the triangular orbit and
including the contributions of the other isolated orbits, describes
very well the coarse-grained quantum-mechanical level density of this
system. We also discuss the role of discrete symmetry for the large 
shell effect obtained here.
\end{abstract}

\maketitle

\section{Introduction}

The periodic orbit theory (POT) is a very useful tool to study shell
structure in single-particle energy spectra. In POT, the
quantum-mechanical level density is semiclassically approximated in
terms of the periodic orbits of the corresponding classical system.
For systems with only isolated orbits, Gutzwiller derived the
so-called ``trace formula'' \cite{Gutz} which is particularly
successful for chaotic systems. The POT for 3D cavities was developed
in \cite{BB}. In integrable systems, semiclassical trace formulae can
be derived from torus quantization \cite{BerryTabor,CrLit1}. However,
most physical systems lie between the above two extreme situations,
i.e., they exhibit mixed phase-space dynamics in which both regular
and chaotic motion coexists on the same energy shell. For systems with
various types of continuous symmetries, trace formulae have been
derived in \cite{CrLit1,StruMa,CrLit2}. For an introductory text book
on semiclassical physics and applications of the POT to various
physical systems, we refer to \cite{BrackText}.

In both integrable and mixed systems, bifurcations of periodic orbits
can significantly influence the shell structure. The above-mentioned
semiclassical trace formulae, which are all based on the
stationary-phase approximation (SPA), diverge when bifurcations of
periodic orbits occur under variations of energy or potential
parameters (e.g., deformations). In the SPA, the classical action
integral is expanded around its stationary points (corresponding to
periodic orbits) up to quadratic terms and the trace integral is
evaluated in terms of Gauss-Fresnel integrals. At bifurcation points,
the determinant of the coefficient matrix of the quadratic terms
becomes zero and the Gauss-Fresnel integral becomes singular. In order
to obtain a finite semiclassical level density near bifurcation
points, higher-order expansion terms of the action integral are
needed, which are most conveniently found from the normal forms
appropriate to the types of bifurcation under consideration
\cite{Ozorio,Sie96,SchSie97}.

Quantum billiards (and their 3D versions, cavities), besides being
quite useful toy models to study POT, reflect important features of
finite physical quantum systems such as quantum dots, metallic
clusters, and atomic nuclei. E.g., non-integrable 3D cavities with
realistic shapes appropriate for fission barriers of actinide nuclei
have been used in POT to explain the onset of the mass asymmetry of
the nascent fission fragments \cite{fisasy}. On the other hand, many
integrable billiard systems are well known and fully understood
semiclassically, e.g., circular, equilateral triangular, square, and
elliptic billiards (cf.\ \cite{BrackText}), and may be used as simple
models for physical systems. In these, bifurcations of short periodic
orbits may lead to a considerable enhancement of shell effects. E.g.,
in the elliptic billiard, the short diametric orbit undergoes
successive bifurcations with increasing deformation, and new
periodic-orbit families emerge \cite{Sie96,Nishioka1,Magner99}. The
same type of bifurcations occur for equatorial orbits in the 3D
spheroidal cavity \cite{Nishioka2} and provide a schematic explanation
of nuclear super-deformed and hyper-deformed shell structure
\cite{ASM,Magner02}. In these studies, it was shown that a system may 
turn strongly chaotic by adding small reflection-asymmetric (e.g., 
octupole) deformations \cite{SAM}.

In this paper, we apply the POT to a two-dimensional non-integrable
billiard system whose boundary is continuously varied from a circular
to an equilateral triangular shape. This study is initiated to explore
a possible quantum dot system. Many studies have been undertaken for
quantum dots, but this type of deformation has not been studied
before. Another important aim is to investigate the role of discrete
symmetries. As will be discussed, the present model possesses the
discrete $C_{3v}$ point-group symmetry. Recently, several mean-field
studies of nuclei have suggested the possible existence of low-lying
states with exotic shapes with point-group symmetries, such as
tetrahedral and octahedral deformations \cite{Dudek}. In order for
such shapes to be stabilized, rather large quantum shell effects in
the single-particle spectra are necessary. We will discuss the role of
discrete symmetries in establishing strong shell effects.

\section{Shell Structure and Level Statistics}

\subsection{The model system}
\begin{figure}[tb]
\includegraphics[width=.75\columnwidth]{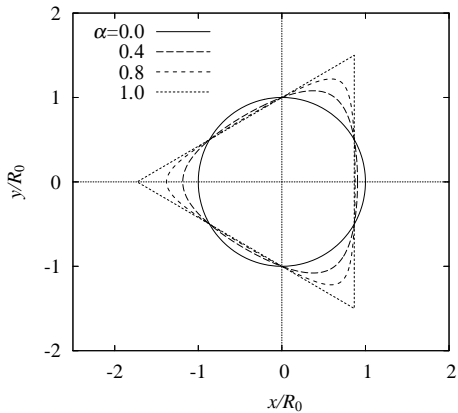}
\caption{\label{fig:shape}
Shapes of the boundary given by Eq.~(\ref{eq:shape}) for several
values of deformation parameter $\alpha$.}
\end{figure}

We consider the two-dimensional billiard system
\begin{equation}
H=\frac{\bm{p}^2}{2m}+V(\bm{r}), \qquad
  V(\bm{r})=\left\{\!\!\begin{array}{c@{\quad}l}
                       0 & r<R(\theta) \\
                       \infty & r>R(\theta)
		       \end{array}\right.
\label{ham}
\end{equation}
in polar coordinates $\bm{r}=(r,\theta)$, whereby the boundary shape 
$R(\theta)$ is parameterized implicitly by the equation
\begin{equation}
R^2+\frac{2\sqrt{3\alpha}}{9}\frac{R^3}{R_0}\cos(3\theta)=R_0^2\,,
\qquad \theta\in [0,2\pi)\,.
\label{eq:shape}
\end{equation}
Figure~\ref{fig:shape} shows the shape of the boundary for several
values of $\alpha$.
This system possesses the discrete symmetries of the point group 
$C_{3v}$, consisting of $\pm2\pi/3$ rotations about the origin and 
reflections with respect to the three axes through the origin with
$\theta=0,\pm\pi/3$. The deformation parameter $\alpha=0$ yields 
a circular shape and $\alpha=1$ an equilateral triangle.
The system is integrable in these two limits, but non-integrable in 
between. 

For the calculation of the quantum spectrum of this system, 
it is useful to decompose the wave functions into free circular waves:
\begin{equation}
\psi_k(r,\theta)
=\sum_{m=-\infty}^\infty c_mJ_{|m|}(kr)e^{im\theta},
\label{eq:cwd}
\end{equation}
where $k=\sqrt{2me}/\hbar$ is the wave number and $e$ the energy.

\begin{figure}[t]
\includegraphics[width=.68\columnwidth]{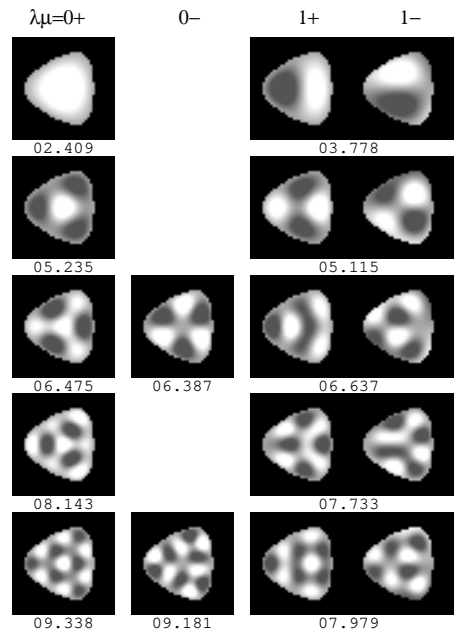}\\
\caption{\label{fig:wfun}
Wave functions of some lowest eigenstates for $\alpha=0.4$.
$0+$, $0-$ and $1\pm$ are states of type
(\ref{eq:wfun1}), (\ref{eq:wfun2}) and (\ref{eq:wfun3}'),
respectively. Their $k$ values are also indicated.}
\end{figure}

\noindent
Taking the $C_{3v}$ symmetry into account, we can classify the eigenstates
according to the eigenvalues of the symmetry operators
\begin{subequations}\label{eq:dsym}

\begin{gather}
\mathcal{R}\psi^{(\lambda\mu)}=e^{2\pi i\lambda/3}\psi^{(\lambda\mu)}, \\
\mathcal{P}\psi^{(\lambda\mu)}=(-1)^\mu\psi^{(\lambda\mu)},
\end{gather}
\end{subequations}
where $\mathcal{R}$ and $\mathcal{P}$ represent rotation by $2\pi/3$
around the origin and reflection with respect to the $x$ axis, respectively.
We have simultaneous eigenstates of $\mathcal{R}$ and $\mathcal{P}$
for $\lambda=0$ states, and then we can classify the eigenstates into 
four sets, which correspond to the irreducible representations
(``irreps'')  of the $C_{3v}$ point group \cite{LandauQM}:
\begin{subequations}\label{eq:wfun}
\begin{gather}
\psi_k^{(0+)}(r,\theta)=\sum_{m=0}^\infty
   c_m^{(0+)} J_{3m}(kr)\cos(3m\theta)\,, \label{eq:wfun1}\\
\psi_k^{(0-)}(r,\theta)=\sum_{m=1}^\infty
   c_m^{(0-)} J_{3m}(kr)\sin(3m\theta)\,, \label{eq:wfun2}\\
\psi_k^{(\pm1)}(r,\theta)=\sum_{m=-\infty}^\infty
   c_m^{(\pm1)}J_{|3m\pm 1|}(kr)e^{i(3m\pm1)\theta}. \label{eq:wfun3}
\end{gather}
\end{subequations}
$\psi_k^{(+1)}$ are the complex conjugates of $\psi_k^{(-1)}$, and the
two form degenerate pairs of states. (The point group
$C_{3v}$ has two 1-dimensional irreps and one 2-dimensional irrep. The
states $(0\pm)$ correspond to the 1-dimensional irreps, and the degenerate
pairs of states $(\pm 1)$ correspond to the 2-dimensional irrep.)
Taking linear combinations of these, one finds the following
alternative real expressions for the states (\ref{eq:wfun3}):
\begin{gather*}
\psi^{(1\pm)}_k(r,\theta)=\!\!\sum_{m=-\infty}^\infty\!\!\!
   c_m^{(1\pm)}J_{|3m+1|}(kr)\left\{
   \begin{array}{c}\cos((3m+1)\theta) \\ \sin((3m+1)\theta)\end{array}
   \right\}\!, \\
\rightline{(\ref{eq:wfun3}')}
\end{gather*}
which are not eigenstates of the operator
$\mathcal{R}$, but of the operator $\mathcal{P}$.
The eigenvalue spectrum $\{k_n\}$ and the coefficients $c_m$ are 
determined by the Dirichlet boundary condition 
$\psi_{k_n}(R,\theta)=0$.
We show some eigenfunctions in Fig.~\ref{fig:wfun}. The states 
belonging to the sets (\ref{eq:wfun1}) and (\ref{eq:wfun2}) have
symmetry under the rotation $\mathcal{R}$. The former are even 
under the reflection $\mathcal{P}$, while the latter are odd.

\begin{figure}[tb]
\includegraphics[width=.75\columnwidth]{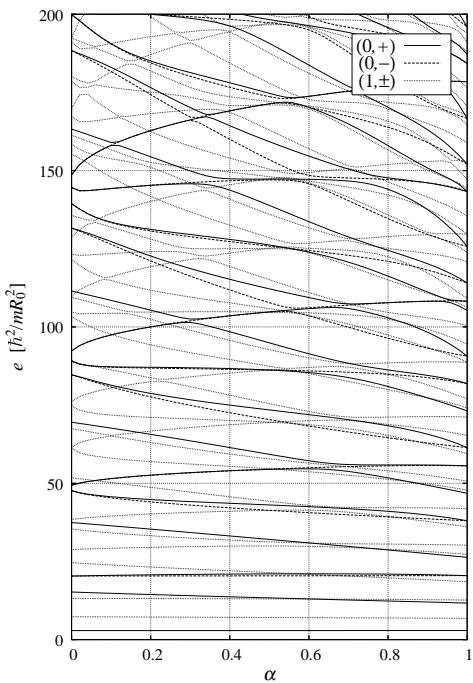}
\caption{\label{fig:sps}
Lowest part of energy spectrum (in units of $\hbar^2\!/mR_0^2$), 
plotted against the deformation parameter $\alpha$.}
\end{figure}

Figure~{\ref{fig:sps}} shows the lower part of the energy 
spectrum $\{e_n\}=\{\hbar^2k_n^2/2m\}$, plotted against the 
deformation parameter $\alpha$.
We note that the levels bunch into small energy intervals in the 
region $\alpha\simeq 0.5 - 0.7$, forming large gaps in the 
spectrum.  It is, in fact, quite surprising to realize that the 
shell gaps in this non-integrable region are much larger than in 
the two integrable limits $\alpha=0$ and $\alpha=1$.
\begin{figure}[tb]
\includegraphics[width=\columnwidth]{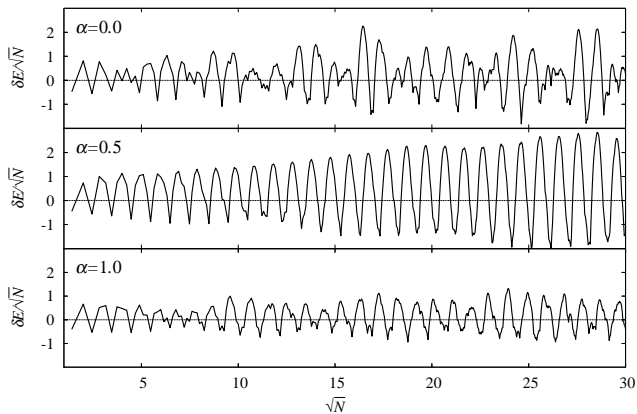}
\caption{\label{fig:sce}
Scaled shell-correction energy $\delta E(N)$
(in units of $\hbar^2/mR_0^2$) versus $\sqrt{N}$.}
\end{figure}
The gaps cause large shell effects in the total
energy of a system of $N$ noninteracting fermions described by
the Hamiltonian (\ref{ham}). We split the energy into a smooth
and an oscillating part
\begin{equation}
E(N)=2\sum_{n=1}^{N/2} e_n = \bar{E}(N)+\delta E(N)\,, \quad (N\hbox{ even})
\label{delE}
\end{equation}
whereby the factor 2 accounts for the spin $s=1/2$.
The smooth part $\bar{E}(N)$ is equivalently given by a Strutinsky
averaging \cite{strut}, the extended Thomas-Fermi model, or the Weyl
expansion (cf.\ \cite{BrackText}, Ch.\ 4), while 
the shell-correction energy $\delta E(N)$
reflects the quantum effects; it 
is dominated by the shortest periodic orbits of the classical system 
as demonstrated below for the case of the level density. Large gaps 
in the spectrum lead to large amplitudes of $\delta E(N)$. This is
demonstrated in Figure\ \ref{fig:sce} where we present $\delta E(N)$,
scaled by a factor $\sqrt{N}$, for three values of $\alpha$. We note
that the oscillatory pattern for $\alpha=0.5$ is quite regular and on
the average has a much larger amplitude than in the integrable limits.

\subsection{Level statistics}

Nearest-neighbor spacing (NNS) distributions are commonly used to
identify signatures of chaos in a quantum system. Generically,
classically chaotic systems exhibit level repulsion and the NNS
distributions are of Wigner type, while regular systems typically have
degeneracies and the NNS distributions are Poisson like
\cite{BoGiSch}.  To extract these universal fluctuation properties,
one has to use \textit{unfolded} spectra whose mean level density is
normalized to unity.  Thus, for systems with discrete
symmetries, one has to study the NNS of the subsets of levels
belonging to the irreps of the corresponding point group.

The average total level density of a two-dimensional billiard is 
given by Weyl's asymptotic formula \cite{Weyl}
\begin{equation}
\bar{g}(e) \sim \frac{m}{2\pi\hbar^2}\,A + {\cal O}(e^{-1/2})\,,
\label{weyl}
\end{equation}
in which the leading-order term is proportional to the area $A$ 
surrounded by the boundary and does not depend on the energy. 
This means that for large energies $e$ the mean level spacing 
$\bar{\Delta}$ becomes asymptotically constant:
\begin{equation}
\bar{\Delta} \; \longrightarrow \; 
\bar{\Delta}_0 = \frac{1}{\bar{g}} \sim \frac{2\pi\hbar^2}{mA}\,.
\end{equation}

As discussed in the previous subsection, the quantum levels of our 
system fall into four sets $\{e_n^{\kappa}\}$ with $\kappa=0\pm$
and $\pm1$, corresponding to the eigenstates 
(\ref{eq:wfun}a,b) and (\ref{eq:wfun3}) or 
(\ref{eq:wfun3}'), respectively. 
The numbers of levels in these sets have the relative ratios
\[
N^{(0+)}:N^{(0-)}:N^{(+1)}:N^{(-1)}=1:1:2:2\,,
\]
and hence the mean level spacing in each set is given by
\begin{equation}
\bar{\Delta}^{(0\pm)}=6\bar{\Delta}_0\,, \qquad
\bar{\Delta}^{(\pm1)}=3\bar{\Delta}_0\,.
\end{equation}
The unfolded level spacings are then obtained by
\begin{equation}
s_n^{(\kappa)} = \frac{e_{n+1}^{(\kappa)}-e_n^{(\kappa)}}
                      {\bar{\Delta}^{(\kappa)}}\,.
                      \qquad (\kappa=0\pm,\pm1)
\end{equation}

\begin{figure}[tb]
\includegraphics[width=\columnwidth]{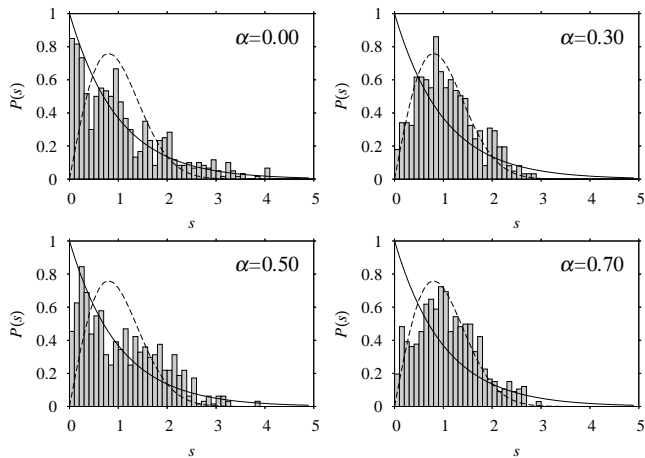}
\caption{\label{fig:nnsd}
Nearest-neighbor spacing distribution $P(s)$ for four values of the
deformation parameter $\alpha$. The lowest 600 levels (i.e., 100 
and 200 levels of the subsets $0\pm$ and $\pm1$, respectively) were
used to obtain the statistics. Solid and dashed lines show
Poisson and Wigner distributions, respectively.}
\end{figure}

\begin{figure}[tb]
\includegraphics[width=.6\columnwidth]{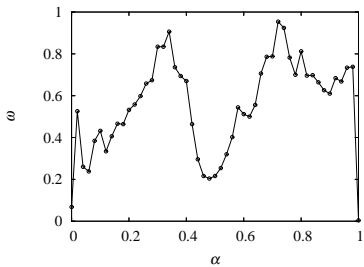}
\caption{\label{fig:brody}
Brody parameter $\omega$ for nearest-neighbor spacing distribution,
plotted as a function of $\alpha$.}
\end{figure}

Figure~\ref{fig:nnsd} shows the NNS distributions $P(s)$, averaged 
over all four sets.  At $\alpha=0$, the system is integrable and the
distribution is Poisson like as expected. At $\alpha=0.3$, the
distribution changes into Wigner form.  A similar situation is also
found at $\alpha=0.7$.  At $\alpha=0.5$, however, the distribution
deviates considerably from the Wigner form and becomes closer to a
Poisson distribution.  For mixed systems, the NNS distributions $P(s)$
are often fitted by a Brody distribution \cite{Brody}
\begin{equation}
B(s,\omega)=\beta(\omega+1)s^\omega\exp(-\beta s^{\omega+1}),
\quad \beta=\left[\Gamma\!\left(\tfrac{\omega+2}{\omega+1}\right)
\right]^{\omega+1}
\label{Brody}
\end{equation}
which interpolates between the Poisson ($\omega=0$) and Wigner 
($\omega=1$) distributions. The Brody parameter $\omega$ can then
be taken as a measure for the chaoticity of NNS distribution.
In Figure \ref{fig:brody} we show $\omega$ as obtained by fitting
the $P(s)$ distributions of Fig.\ \ref{fig:nnsd} to (\ref{Brody}).
We clearly recognize two peaks around $\alpha\simeq 0.3$ and $\simeq 0.7$,
exhibiting near-chaoticity, separated by a deep minimum around
$\alpha\simeq 0.5$ where the system appears to approach regularity.
As we will discuss below, this near-regularity is related to an 
approximate restoration of local dynamical symmetry due to a 
periodic-orbit bifurcation.

\section{Fourier Analysis and Classical Periodic Orbits}
\label{sec:orbits}

In the POT, the quantum level density $g(e)$ is approximated
in terms of classical periodic orbits by the semiclassical trace 
formula \cite{Gutz,BB,BerryTabor,StruMa,CrLit1,CrLit2}
\begin{align}
g(e) & = \sum_n \delta(e-e_n)\nonumber\\
     & \simeq  \bar{g}(e)+\sum_\xi A_\xi(e)\cos\left(\frac{S_\xi(e)}{\hbar}
                -\frac{\pi}{2}\nu_\xi\right)\!,~~ \label{eq:trace_e}
\end{align}
where the first term, like for the energy in (\ref{delE}), represents
the smooth part, while the second term contains the quantum 
shell effects. In the latter, the sum is taken over all periodic
orbits $\xi$ of the classical system (or the orbit families $\xi$ in 
systems with continuous symmetries \cite{BerryTabor,StruMa,CrLit1}), $S_\xi$ 
is the action integral around $\xi$, and $\nu_\xi$ is the Maslov index 
\cite{CrRoLi,Sugita}. The amplitude $A_\xi$ depends on the stability
of the orbit $\xi$ (and, for an orbit family, on the phase-space
volume covered by the family). For isolated orbits, the amplitude 
$A_\xi$ was given by Gutzwiller \cite{Gutz}:
\begin{equation}
A_\xi(e) = \frac{1}{\pi\hbar}\,\frac{T_\xi(e)}{\sqrt{|\det[1-M_\xi(e)]|}},
\label{ampgutz}
\end{equation}
where $T_\xi(e)={\rm d}S_\xi(e)/{\rm d}e$ is the period of the orbit 
and $M_\xi(e)$ its stability matrix defined below. If an orbit has 
a discrete degeneracy $f$ (i.e., if there exist $f$ replicas with
identical actions, stabilities and Maslov indices but different
orientations) due to discrete symmetries, it has to be included
$f$ times into the sum in (\ref{eq:trace_e}). This holds also for
time-reversed rotational orbits.

Transforming variables from energy $e$ to wave number $k$ and using the
relation $S_\xi=\hbar kL_\xi$ for billiards, where $L_\xi$ is the
length of the orbit $\xi$, the trace formula becomes
\begin{align}
g(k) & = \frac{\hbar^2k}{m}g(e)\nonumber\\
     & \simeq \bar{g}(k)+\sum_\xi A_\xi(k)\cos(kL_\xi-\frac{\pi}{2}\nu_\xi)\,.
\label{eq:trace_k}
\end{align}
Let us now consider the Fourier transform of the level density $g(k)$ 
with respect to $k$:
\begin{equation}
F(L)=\int_{-\infty}^\infty g(k)e^{-ikL}e^{-(k\Delta)^2/2}{\rm d}k.
\end{equation}
The Gaussian damping factor is included for the truncation of the 
high-energy part of the spectrum. If we insert Eq.~(\ref{eq:trace_k}) 
and ignore the $k$ dependence of the amplitude factors $A_\xi$ (which 
are constant for isolated orbits in billiards 
and cavities), we obtain
\begin{equation}
F^{\rm sc}(L)=F_0(L)+\pi\sum_\xi e^{-i\pi\nu_\xi/2}A_\xi\,
  \delta_\Delta(L-L_\xi)\,. \label{eq:fourier_cl}
\end{equation}
$\delta_\Delta(x)$ is a normalized Gaussian of width $\Delta$, which
turns into Dirac's delta function in the limit $\Delta\to 0$.
Eq.~(\ref{eq:fourier_cl}) indicates that the Fourier transform
$F(L)$ is a function with successive peaks at the lengths of the periodic
orbits $L=L_\xi$, with heights proportional to the amplitudes
$A_\xi$. We can therefore extract information about the classical
periodic orbits from the Fourier transform of the quantum-mechanical 
level density:
\begin{equation}
F^{\rm qm}(L) = \sum_n e^{-ik_nL-(k_n\Delta)^2}\!, \quad
      k_n = \sqrt{2me_n}/\hbar\,.
\label{eq:fourier_qm}
\end{equation}

\begin{figure}
\begin{flushleft}
\includegraphics[width=\columnwidth]{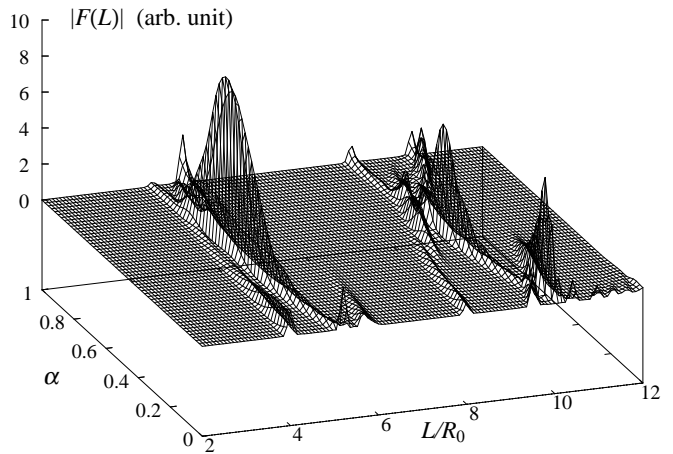}
\end{flushleft}\vspace*{-\baselineskip}
\caption{\label{fig:ftl3d}
Modulus $|F^{\rm qm}(L)|$ of Fourier transform of the quantum-mechanical 
level density versus deformation parameter $\alpha$.}
\end{figure}

Figure~\ref{fig:ftl3d} shows the modulus of this Fourier transform
of our system versus the deformation parameter $\alpha$. At
$\alpha=0$, where the periodic orbits form degenerate families 
corresponding to rational tori, we find the peaks at the well-known 
orbit lengths of the circular billiard \cite{BB}
\begin{equation}
L_{vw}=2vR_0\sin(w\pi/v)\,.
\end{equation}
Here $w$ and $v$ are the winding number around the origin and the
number of vertices $(v\geq 2w)$ of each orbit, respectively.  For
instance, $L=4R_0$ for the diametric orbit $(w=1,v=2)$,
$L=3\sqrt{3}R_0$ for the triangular orbit $(w=1,v=3)$,
$L=4\sqrt{2}R_0$ for the square orbit $(w=1,v=4)$, and so on. With
increasing $\alpha$, we observe a dramatic enhancement of the peak
height corresponding to the triangular orbit, $L\simeq5.2R_0$ (and its
second repetition, $L\simeq10.4$), starting around $\alpha\simeq 0.4$
and culminating around $\alpha\simeq0.55$.  As we shall see below,
this can be traced back to a codimension-two bifurcation of the
triangular orbit.

In fact, for $\alpha>0$ all rational tori of the circular billiard 
are broken into pairs of stable and unstable isolated orbits. With 
increasing $\alpha$, bifurcations of the stable orbits occur and 
new periodic orbits emerge, which makes the phase-space increasingly 
chaotic. In the following, we first demonstrate this for the two 
shortest orbits and then focus on the triangular orbit. 
 
The stability of a periodic orbit is described by the stability matrix 
$M_\xi$, which is defined by the linearized Poincar\'{e} map around 
the periodic orbit
\begin{equation}
M_\xi
=\pp{(\bm{q}(T_\xi),\bm{p}(T_\xi))}{(\bm{q}(0),\bm{p}(0))},
\end{equation}
where $(\bm{q}(t),\bm{p}(t))$ are the coordinates and momenta
perpendicular to the periodic orbit $\xi$ as functions of time 
$t$, and $T_\xi$ is the period of the orbit. In a two-dimensional 
autonomous Hamiltonian system $M_\xi$ is a symplectic (2x2)-matrix, 
and the stability of a orbit is easily identified by looking at 
its trace $\Tr M_\xi$. For elliptic (stable) orbits, 
the eigenvalues of $M_\xi$ are of the form $(e^{iv},e^{-iv})$ 
with real $v\geq0$, and thus $-2\leq\Tr M_\xi\leq 2$. For direct-
and inverse-hyperbolic (unstable) orbits, the eigenvalues are 
$(e^u, e^{-u})$ and $(-e^u,-e^{-u})$, respectively, with real $u>0$, 
and hence $\Tr M_\xi > 2$ and $\Tr M_\xi < -2$. Bifurcations of
isolated orbits occur whenever $\Tr M_\xi = +2$.

\begin{figure}[t]
\includegraphics[width=.8\columnwidth]{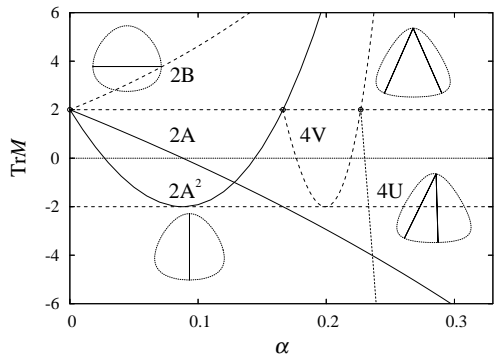}
\caption{\label{fig:trm2}
Trace of stability matrix $\Tr M_\xi$ for the diameter orbits (2A)
and the second repetition of the stable diameter orbit (2A$^2$),
plotted versus deformation parameter $\alpha$.}
\end{figure}

\begin{figure}[t]
\includegraphics[width=.85\columnwidth]{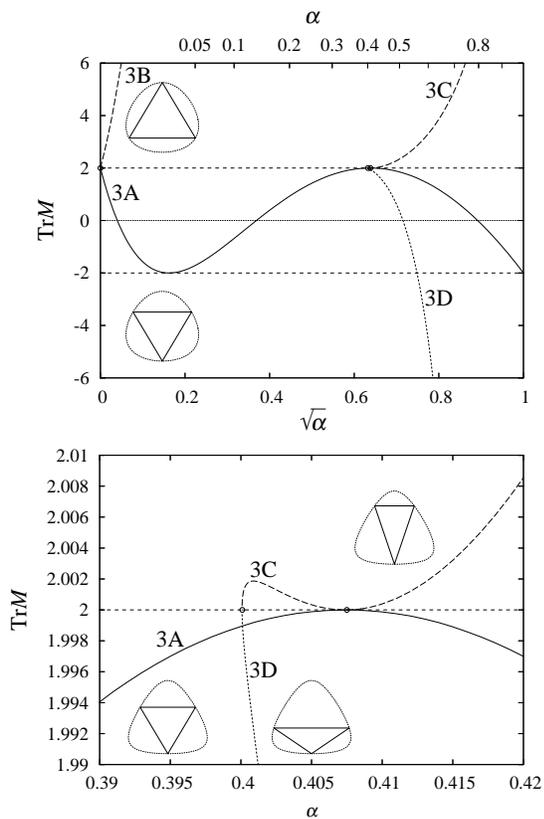}
\caption{\label{fig:trm3}
Same as Fig.~\ref{fig:trm2} but for the triangular orbits,
in the upper panel plotted versus $\sqrt{\alpha}$.
The lower panel shows a magnification around the bifurcation
of the stable orbit 3A occurring near $\alpha\simeq0.407$.}
\end{figure}

Figures~\ref{fig:trm2} and \ref{fig:trm3} show $\Tr M_\xi(\alpha)$
for the two shortest pairs of periodic orbits in our system. 
In Fig.~\ref{fig:trm2}, the stable branch (2A) of the diametric 
orbit is seen to undergo a period-doubling pitchfork bifurcation at 
$\alpha=0.166$, where a symmetric wedge-shaped orbit (4V) emerges. 
The latter undergoes a further pitchfork bifurcation at 
$\alpha=0.227$, where a pair of asymmetric wedge-shaped orbits 
(4U) emerges.

The bifurcation scenario of the triangular orbits is more complicated. 
In the upper panel of Fig.~\ref{fig:trm3}, where $\Tr M_\xi$ is 
plotted against $\sqrt{\alpha}$, the stable triangular orbit (3A) 
is seen to touch the critical line $\Tr M_\xi=+2$ near 
$\sqrt{\alpha}\simeq0.63$ ($\alpha\simeq0.4$) and to remain
stable on either side. A pair of new stable (3D) and unstable 
(3C) triangular orbits emerge from the touching point. A magnification 
of the situation around $\alpha\simeq0.4$, shown in the lower panel
against $\alpha$, reveals that the scenario consists of two connected 
near-lying bifurcations, also called a ``codimension-two bifurcation''. 
At $\alpha=0.400$, a tangent (saddle-node) bifurcation occurs, at 
which the new orbits 3D and 3C are born. Shortly after this bifurcation, 
the (old) stable orbit 3A and the new unstable orbit 3C encounter
in a touch-and-go bifurcation at $\alpha=0.407$ (see
Appendix~\ref{app:tracem} for its analytic value).  Note that the 
new 3C and 3D orbits do not possess $C_{3v}$ symmetry, in contrast to 
the old 3A orbit, so that they occur in degenerate triplets obtained 
by successive rotations about $2\pi/3$, i.e., these orbits have
a discrete degeneracy of $f=3$.

\begin{figure*}
\includegraphics[width=\textwidth]{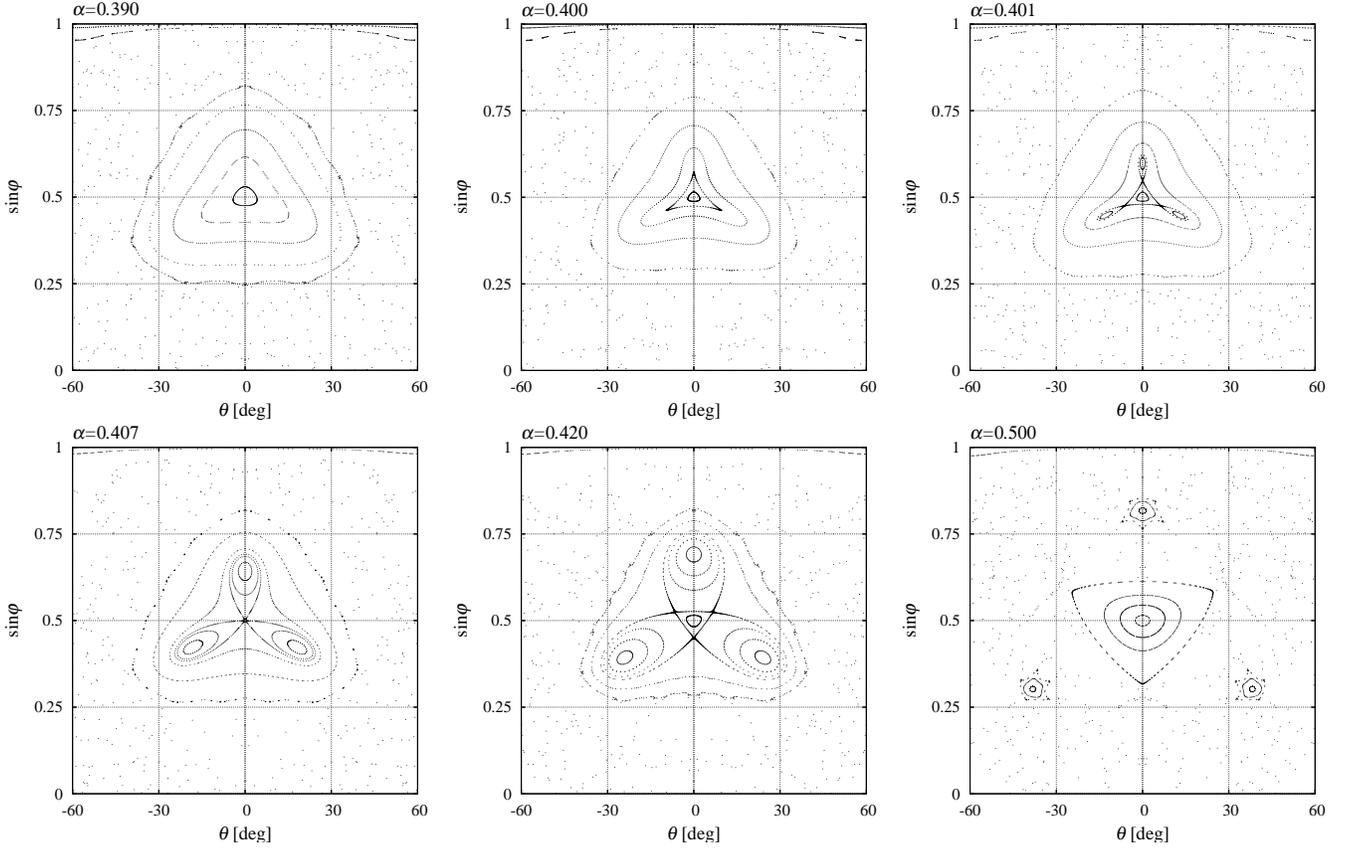}
\caption{\label{fig:pmap}
Poincar\'e surfaces of section $(\theta,\sin\varphi)$ in the
fundamental domain of the $C_3$ group, plotted for several values of
$\alpha$ (given at the upper left of each panel) in the bifurcation
region of the triangular orbit 3A. $\theta$ is the polar angle of a
reflection point and $\varphi$ the reflection angle measured from the
normal to the boundary; the set $(\theta,\sin\varphi)$ is 
approximately area preserving.}
\end{figure*}

Figure~\ref{fig:pmap} shows excerpts of Poincar\'e surfaces of section
$(\theta,\varphi)$ in the relevant regions. Here $\theta$ is the
polar angle of a reflection point of an orbit at the boundary,
while $\varphi$ represents the reflection angle measured from the
normal to the boundary at the reflection point. The three upper
panels illustrate the tangent bifurcation of the 3C and 3D orbits. 
For $\alpha=0.390$ (upper left), one sees one major regular island
corresponding to the stable equilateral triangular orbit 3A.
It contains its fixed point at $(\theta,\varphi)=(0,\pi/6)$,
surrounded by quasi-tori (small aperiodic perturbations of the 3A orbit).
At the bifurcation point $\alpha\approx 0.400$ 
(upper center), three cusps are formed by one of the surrounding 
quasi-tori in the island, and for $\alpha=0.401$ (upper right), the
three stable fixed points of the new 3D orbits, surrounded by small
regular islands, are seen to have emerged from each of the three 
cusps in the major island. The three saddles separating
these three islands from the central island contain the 
unstable fixed points of the 3C orbits.
The stable fixed point of the 3A orbit still persists at the original
position at the center of the major island. The number $f=3$ of the
new stable and unstable fixed points in the island is due to the
three-fold discrete degeneracy of the orbits 3D and 3C mentioned
above.

The lower three panels in Fig.~\ref{fig:pmap} illustrate the
touch-and-go bifurcation of the orbits 3A and 3C. At the bifurcation
point $\alpha=0.407$ (lower left),
the three unstable fixed points of the 3C orbits
have contracted into a star-like intersection of three quasi-tori,
located at the central fixed point $(\theta,\varphi)=(0,\pi/6)$ of
the 3A orbit.  The three nearby stable fixed points in the major
island belong to the orbits 3D.  For $\alpha=0.420$ (lower center),
small stable islands have formed again around the fixed points of the
3A orbit, and nearby we recognize the three unstable fixed points of
the 3C orbit.  For $\alpha=0.500$ (lower right), the central island of
the 3A orbits have grown, the three islands of the stable 3D orbits
have shrunk, and the three unstable fixed points of the 3C orbits
are about to be buried in the increasing chaotic structure.

The bifurcation of the triangular orbit 3A occurs at the deformation 
where we observed the onset of the large Fourier 
peak in Fig.~\ref{fig:ftl3d}. It is due to the appearance of the pair 
of six-fold degenerate new orbits 3D and 3C. In particular, the stable
one (3D), besides the doubly-degenerate 3A orbit, adds to the local
regularity of the phase space observed in Figs.~\ref{fig:nnsd},
\ref{fig:ftl3d} and \ref{fig:pmap}. It is therefore apparent that 
this bifurcation is responsible for the remarkable shell structure 
in the quantum spectrum seen in Fig.~\ref{fig:sps}. 
Note that the maximum of the Fourier peak in
Fig.~\ref{fig:ftl3d} occurs at $\alpha\simeq0.55$, i.e., above the
bifurcation where the new orbits are born. The fact that the 
signature of a bifurcation is strongest {\it after} the bifurcation 
point (on the side where the new orbits exist) is a rather general 
trend for shell effects which can also be understood in terms of 
semiclassical uniform approximations \cite{Sie96,SchSie97} (see the 
following section and Fig.~\ref{fig:uaamp3} below). It has also been 
observed recently in the level statistics of a Hamiltonian system with 
mixed phase space \cite{Marta}.

\section{Semiclassical Analysis}

\subsection{Semiclassical level density near bifurcations}

As mentioned above, the Gutzwiller trace formula (\ref{eq:trace_e}) with the
amplitudes (\ref{ampgutz}) for isolated orbits diverges at bifurcation
points. This is due to the break-down of stationary phase approximation, 
and higher-order terms in the expansion of the action integral must be 
included in the derivation of the trace formula. For this purpose, it
is necessary \cite{Ozorio} to express the trace integral
in phase space. After integrating over the coordinate and momentum
parallel to an isolated orbit $\xi$ in a two-dimensional system, its 
semiclassical contribution to the oscillating part of the level
density becomes \cite{CrLit1,SchSie97,Magner02}
\begin{align}
\delta g_\xi(e)=&\frac{1}{2\pi^2\hbar^2}\Re\int_{-\infty}^\infty {\rm d}q'
  \int_{-\infty}^\infty {\rm d}p\,
  \frac{1}{n_r}\,\frac{\partial\widehat{S}}{\partial e}\,
  \left|\pp{^2\widehat{S}}{p\partial q'}\right|^{1/2} \nonumber\\
  &\times\exp\!\left[\frac{i}{\hbar}\widehat{S}(q',p)
  -\frac{i}{\hbar}q'p-\frac{i\pi}{2}\nu_\xi\right].
\label{eq:phtrace}
\end{align}
Here $q'$ is the final coordinate and $p$ the initial momentum
perpendicular to the orbit $\xi$, the two forming a canonical pair
of variables $(q',p)$ to describe the ``natural'' Poincar\'e surface 
of section (PSS) of the orbit on which $(q',p)=(0,0)$ is its fixed point.
For generic bifurcations, $n_r$ is the repetition number of the 
primitive orbit. The function $\widehat{S}(q',p)$ denotes the Legendre 
transform of $S(q',q)$:
\begin{equation}
\widehat{S}(q',p)=S(q',q)+q'p\,,
\label{Leg}
\end{equation}
where $S(q',q)$ is the (open) action integral along the orbit $\xi$
(at fixed energy $e$)
\begin{equation}
S({\bm{r'},\bm{r}) = \int_{\bm{r'}}^{\bm{r}} {\bm p}_\xi(\bm r''})
                     \cdot{\rm d}\bm{r''},
\end{equation}
projected on the $q$ axis. The function $\widehat{S}(q',p)$ is actually the
generating function of the Poincar\'e map:
\begin{equation}
\widehat{S}(q',p): \quad (q,p)\rightarrow(q',p')\,.
\end{equation}

If one expands the function $\Phi(q',p)=\widehat{S}(q',p)-q'p$
in the phase of the integrand of (\ref{eq:phtrace}) around $q'=p=0$ 
up to quadratic terms in $q'$ and $p$ and evaluates the integrals in 
(\ref{eq:phtrace}) by the standard stationary phase approximation (SPA), 
one obtains the contribution of the orbit $\xi$ to Gutzwiller's trace 
formula (\ref{eq:trace_e}) with the amplitude (\ref{ampgutz}). Near bifurcations
of the orbit $\xi$, the SPA breaks down and one has to include 
higher-order terms in the expansion of $\Phi(q',p)$. The minimum
number of terms needed to describe a given bifurcation scenario on 
the PSS leads to the so-called ``normal forms'' of $\Phi(q',p)$,
which depend on the type of bifurcation. Doing the integrations in 
(\ref{eq:phtrace}) using such normal forms leads to a finite combined 
contribution of the orbit $\xi$ and all the orbits involved in its 
bifurcation to the trace formula. In order to conform with standard
notation, we rename the function $\Phi(q',p)$ by $S(q',p)$ which,
according to (\ref{Leg}), is identical with the projected action
integral $S(q',q)$, but taken as a function of the variables $q'$ and $p$. 

\subsection{Normal form for codimension-two bifurcation}

For the description of the codimension-two bifurcation scenario of 
the orbit 3A and its satellites 3C, 3D discussed in the previous 
section and illustrated on the PSS in Fig.~\ref{fig:pmap}, the 
following normal form is appropriate \cite{Schom97}:
\begin{equation}
S(I,\varphi)=S_0(\alpha)-\epsilon I-aI^{3/2}\cos(3\varphi)-bI^2\,.
\label{eq:normalform}
\end{equation}
Here one has transformed the Poincar\'e variables $(q',p)$ to
quasi-polar variables $(I,\varphi)$ by
\begin{equation}
p=\sqrt{2I}\cos\varphi\,,\qquad q'=\sqrt{2I}\sin\varphi\,.
\end{equation}
In (\ref{eq:normalform}), $S_0(\alpha)$ is the (closed) action 
integral of the central (3A) orbit as a function of $\alpha$. The
``bifurcation parameter'' $\epsilon$ is a monotonously decreasing 
function of $\alpha$ such that $\epsilon=0$ at the touch-and-go 
bifurcation point (here $\alpha=\alpha_0=0.407$) of the central 
orbit, and that $\epsilon<0$ for $\alpha>\alpha_0$. $a$ and $b$ 
are parameters which are specific for the system and will be 
determined below.

That (\ref{eq:normalform}) is able to describe the correct 
fixed-point structure not only of the 3A orbit but also of its 
satellites 3C and 3D, will now be shown explicitly. We first rewrite
(\ref{eq:normalform}) in terms of $q'$ and $p$:
\begin{equation}
S(q',p)=S_0-\frac{\epsilon}{2}(p^2+q'^2)
  -\frac{a}{\sqrt{8}}(p^3-3pq'^2)-\frac{b}{4}(p^2+q'^2)^2.
\label{eq:action}
\end{equation}
The stationary-phase conditions are
\begin{gather}
\pp{S}{q'}\Bigr|_{q_i}=0\,, \qquad \pp{S}{p}\Bigr|_{p_i}=0\,.
\end{gather}
One of the solutions is $q_0=0$, $p_0=0$ and corresponds to the central
3A orbit. This is so by default, due to the choice of the Poincar\'e
variables $(q,p)$. Two further stationary points are found to be
\begin{equation}
q_{1,2}=0\,, \qquad
p_{1,2}=-\frac{3a}{4\sqrt{2}\,b} \pm\sqrt{\frac{9a^2}{32b^2}-\frac{\epsilon}{b}}\,.
\label{qp12}
\end{equation}
The fixed points $(q_i,p_i)$ with $i=1,2$ belong to the satellite
orbits 3C and 3D, respectively. Two more pairs of fixed points for
each of them are found by rotations in the $(I,\varphi)$ plane:
$\varphi\to\varphi\pm 2\pi/3$. $\epsilon_1=9a^2\!/32b$ is the
critical value for $p_{1,2}$ to have real values, i.e., for the real
orbits 3C and 3D to exist. For our system we can choose $b>0$, and
therefore $\epsilon_1>0$. For $\epsilon>\epsilon_1$, the orbits 3C and
3D become imaginary and only the central orbit 3A is real. With
decreasing $\epsilon$, the three pairs of stable and unstable
satellite orbits appear at $\epsilon=\epsilon_1$ which we identify
with the tangent bifurcation point. At $\epsilon=0$ we have the
touch-and-go bifurcation as stated above. For $\epsilon<0$
($\alpha>\alpha_0$), all orbits are real.

The normal-form parameters $\epsilon$, $a$ and $b$ can be determined
by fitting the actions $S$ and stability traces $\Tr M$ of the orbits,
which have been obtained numerically, to their local behaviors
predicted by the normal form (\ref{eq:normalform}). The stability
trace is given in terms of $\widehat{S}$ by \cite{SchSie97}
\begin{equation}
\Tr M=\left(\!\pp{^2\widehat{S}}{p\partial q'}\!\right)^{\!\!-1}\!
\left[1+\left(\!\pp{^2\widehat{S}}{p\partial q'}\!\right)^{\!\!2}\!\!
-\pp{^2\widehat{S}}{p^2}\pp{^2\widehat{S}}{{q'}^2}\right]\!\!.
\end{equation}
For the central orbit, it becomes
\begin{equation}
\Tr M_0=\Tr M_A=2-\epsilon^2\,,
\end{equation}
so that $\epsilon$ can be determined as $\epsilon=\pm\sqrt{2-\Tr
M_A}$, choosing the correct sign on either side of the bifurcation. 
The other parameters are obtained from the action difference of the 
satellite orbits 1 and 2 (i.e., 3C and 3D). Inserting 
$(q_{1,2},p_{1,2})$ from (\ref{qp12}) into (\ref{eq:action}), one finds
\begin{gather}
\frac{S_D-S_C}{\hbar kR_0}
=\frac{4}{3b}\sqrt{\epsilon_1}(\epsilon-\epsilon_1)^{3/2}.
\end{gather}
By fitting the numerical data for $S_D-S_C$ as function of $\epsilon$,
we obtain $\epsilon_1$ and $b$, and therefore $a$. Thus we can uniquely 
determine all the normal form parameters. The results for the
triangular orbits discussed above are
\begin{equation}
a=\frac{0.519252}{\sqrt{\hbar kR_0}}\,, \;\,
b=\frac{2.34950}{\hbar kR_0}\,,  \;\,
\epsilon_1=0.0322755\,. 
\label{eq:nparam3}
\end{equation}
The formulae for the stability traces for the orbits 3C and 3D are
\begin{align}
\Tr M_{C,D} & =  2\pm12\sqrt{\epsilon_1^3(\epsilon_1-\epsilon)}
                  -24\epsilon_1(\epsilon_1-\epsilon) \nonumber\\
            &    \quad \pm12\sqrt{\epsilon_1(\epsilon_1-\epsilon)^3}\,,
\end{align}
and the action difference between orbits 3C and 3A is
\begin{equation}
\frac{S_C-S_A}{\hbar kR_0}
=\frac{1}{12b}\!\left[\,3\epsilon^2\!-12\epsilon_1\epsilon
+8\epsilon_1^2\!-8\sqrt{\epsilon_1(\epsilon_1-\epsilon)^3}\,\right]\!.
\end{equation}
We have checked that the numerical results in the neighborhood of the
bifurcations are nicely reproduced by these equations.

\subsection{Uniform approximations}
Inserting the normal form (\ref{eq:normalform}) into the integral 
(\ref{eq:phtrace}), one obtains a ``local'' uniform approximation 
\cite{Ozorio} which is finite and valid near the bifurcation, i.e., 
for not too large absolute values of $\epsilon$. Due to the $C_{3v}$ 
symmetry of our system, the touch-and-go bifurcation is non-generic
and isochronous. The factor $n_r$ in the denominator of 
(\ref{eq:phtrace}) here must be chosen \cite{note} as $n_r=f=3$.
We have another degeneracy factor 2 in the numerator due to the 
time-reversal symmetry of all orbits. Using the 
variables $(I,\varphi)$ and changing the energy $e$ to the wave 
number $k$, we finally obtain the following expression for the
local uniform approximation, in which the integration over $\varphi$
can be done analytically:
\begin{align}
&\delta g_\xi(k)=\frac{1}{3\pi^2\hbar}\, \Re
  \int_0^\infty {\rm d}I\int_0^{2\pi}{\rm d}\varphi\,
  L_\xi\left|\pp{^2\widehat{S}}{I\partial\varphi}\right|^{1/2}
  \nonumber\\
  &\times\exp\left[\frac{i}{\hbar}\!\left\{
  S_0-I\varphi-\epsilon I-aI^{3/2}\cos(3\varphi)-bI^2\right\}
  -\frac{i\pi}{2}\nu_\xi\right] \nonumber\\
  &=\frac{2L_\xi}{3\pi\hbar}\Re e^{ikL_\xi-i\pi\nu_\xi/2}
   \!\!\int_0^\infty\!\! {\rm d}I\,J_0(\tfrac{a}{\hbar}I^{3/2})
  e^{\frac{i}{\hbar}(-\epsilon I-bI^2)}.
\label{locunif}
\end{align}
The integration over $I$ can be performed numerically using an expansion
formula given in \cite{Schom97}.

As stated above, the result (\ref{locunif}) is only useful in the
neighborhood of the bifurcation. Far away from it, where all involved
orbits become isolated, it can be evaluated asymptotically
(corresponding to the SPA), but the amplitudes of the orbits then do 
not agree with the Gutzwiller values (\ref{ampgutz}). In order to 
achieve this, one must develop ``global'' uniform approximations
\cite{Sie96,SchSie97}. To that purpose, one needs to include
higher-order expansion terms in the normal form. For the 
codimension-two bifurcation of our type one obtains, after suitable 
coordinate transformations \cite{Schom98},
\begin{align}
&\delta g_\xi(k)=\frac{1}{3\pi^2\hbar}\Re
\int_0^\infty {\rm d}I\int_0^{2\pi}{\rm d}\varphi\,\Psi(I,\varphi) \nonumber\\
&\times\exp\left[\frac{i}{\hbar}\left\{S_0-I\varphi\!-\epsilon I
-aI^{3/2}\cos(3\varphi)-bI^2\right\}-\frac{i\pi}{2}\nu_\xi\right]
\label{globunif}
\end{align}
with
\begin{equation}
\Psi(I,\varphi)=L_\xi-\alpha I-\beta I^{3/2}\cos(3\varphi),
\end{equation}
where $\alpha$ and $\beta$ are expressed by a certain combination of
higher-order expansion coefficients. In practice, these parameters
are determined such that (\ref{globunif}) yields the sum of 
contributions with amplitudes (\ref{ampgutz}) of all involved
isolated orbits far away from the bifurcation point. They are 
determined by equating (cf.\ \cite{Kaidel2})
\begin{gather}
\frac{L_C}{\sqrt{|2-\Tr M_C|}}
=\frac{L_A-\alpha I_C-\beta I_C^{3/2}}{\sqrt{|\det\Phi''(I_C)|}}\,,\\
\frac{L_D}{\sqrt{|2-\Tr M_D|}}
=\frac{L_A-\alpha I_D-\beta I_D^{3/2}}{\sqrt{|\det\Phi''(I_D)|}}\,,
\end{gather}
with 
\begin{equation}
\det\Phi''=\begin{vmatrix}
  \pp{^2S}{{\varphi}^2} & \pp{^2S}{\varphi\partial I} \\
  \pp{^2S}{I\partial\varphi} & \pp{^2S}{I^2}
  \end{vmatrix}_{\varphi=0}
 =\begin{vmatrix}
  \pp{^2S}{{q'}^2} & \pp{^2S}{q'\partial p} \\
  \pp{^2S}{p\partial q'} & \pp{^2S}{p^2}
  \end{vmatrix}_{q'=0}\!\!\!,
\end{equation}
and
\begin{equation}
I_{C,D} = \frac12\left(q_{1,2}^2+p_{1,2}^2\right).
\end{equation}

We now have all ingredients ready for calculating the semiclassical 
level density of our system. Since it is well-known \cite{Gutz} that
the sum over all periodic orbits does not converge in systems with
mixed dynamics, we have to introduce a certain truncation. This is
achieved \cite{BrackText,SieSt} by focusing on the gross-shell structure 
of the level density, coarse-graining it by a Gaussian convolution 
over $k$ with width $\gamma$:
\begin{equation}
g_\gamma(k)=\frac{1}{\sqrt{\pi}\gamma}\int {\rm d}k'\,e^{-[(k'-k)/\gamma]^2}g(k')\,.
\end{equation}
The semiclassical trace formula (\ref{eq:trace_k}) then becomes
\begin{equation}
g_\gamma(k)\simeq\bar{g}(k)+\sum_\xi e^{-(\gamma L_\xi/2)^2}
  A_\xi(k)\cos(kL_\xi-\frac{\pi}{2}\nu_\xi).
\label{coarsetf}
\end{equation}
Due to the Gaussian factor, periodic orbits with lengths $L_\xi\gtrsim
2\pi/\gamma$ are now exponentially suppressed. Choosing $\gamma=0.6$, we
need only consider periodic orbits with $L_\xi < 10R_0$. From 
Fig.~\ref{fig:trm2}, we see that no isochronous bifurcations of the 
diametric orbit occur for $\alpha>0$, so that its contribution can be 
evaluated by the standard Gutzwiller formula with amplitudes 
(\ref{ampgutz}). A period-doubling bifurcation of the stable diameter 
occurs at $\alpha\simeq 0.15$, but does not contribute much to the 
coarse-grained level density. The same is true for the tetragonal and 
pentagonal orbits. The contributions of the triangular orbit 3A and 
its satellites 3C, 3D are included in the global uniform approximation 
(\ref{globunif}), including the Gaussian damping factor 
$\exp[-(\gamma L_{A}/2)^2]$.

\begin{figure}[tb]
\includegraphics[width=.8\columnwidth]{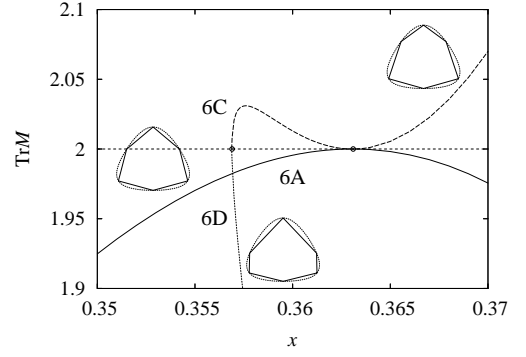}
\caption{\label{fig:trm6}
Same as Fig.~\ref{fig:trm3} for the hexagonal orbits.}
\end{figure}

\begin{figure}[tb]
\includegraphics[width=.49\columnwidth]{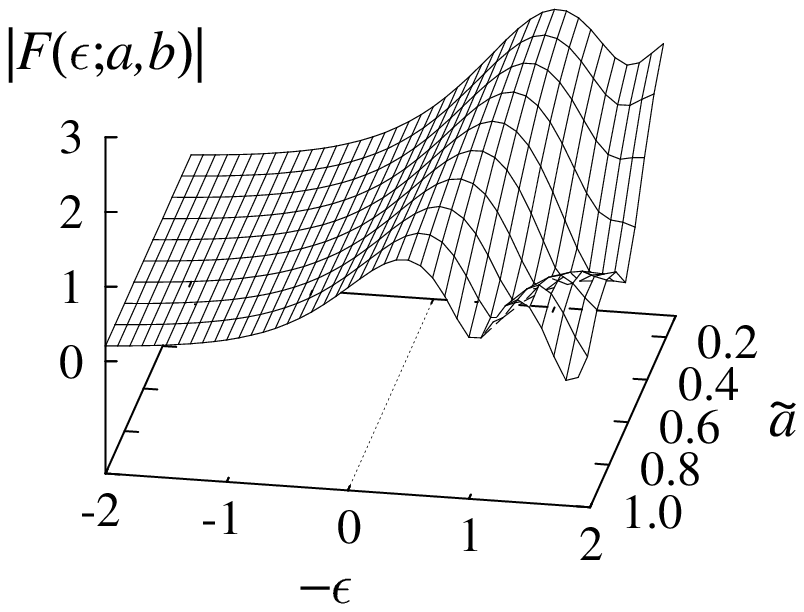} \hfill
\includegraphics[width=.49\columnwidth]{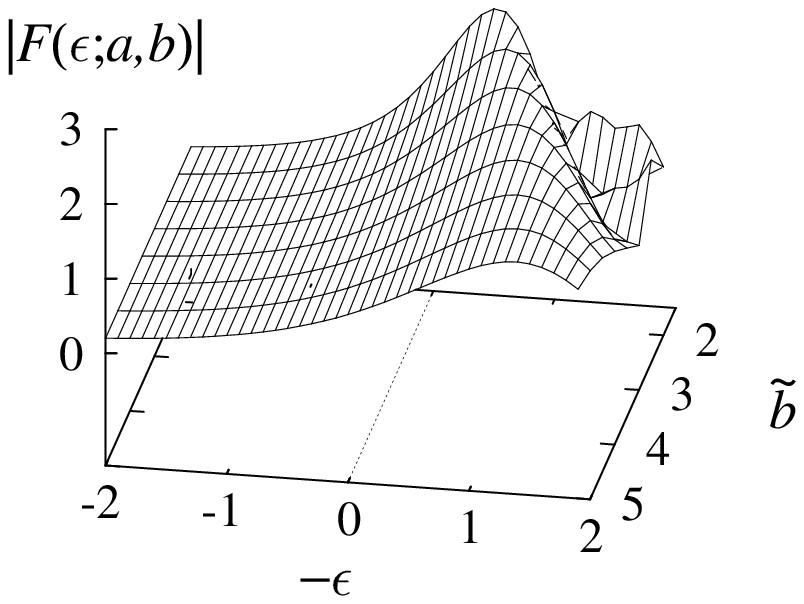} \\
\caption{\label{fig:uaamp3}
Profile of the function $F(\epsilon;a,b)$ in
(\ref{eq:uaamp}) for the normal form parameters (\ref{eq:nparam3})
of the triangular orbit. The left panel is for fixed $b$ and varying
$\tilde{a}=a\sqrt{\hbar kR_0}$, and the right panel for fixed $a$ and
varying $\tilde{b}=b\hbar kR_0$, both for $kR_0=20$.}
\medskip

\includegraphics[width=.49\columnwidth]{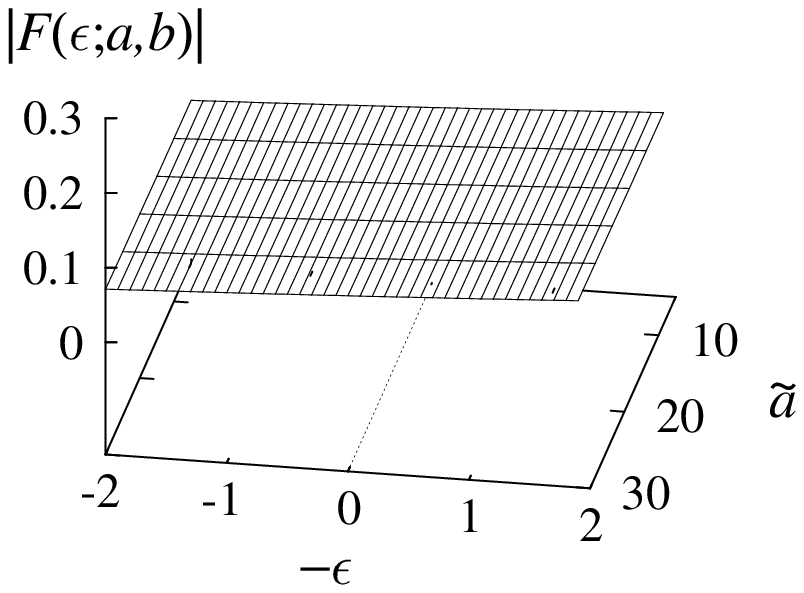} \hfill
\includegraphics[width=.49\columnwidth]{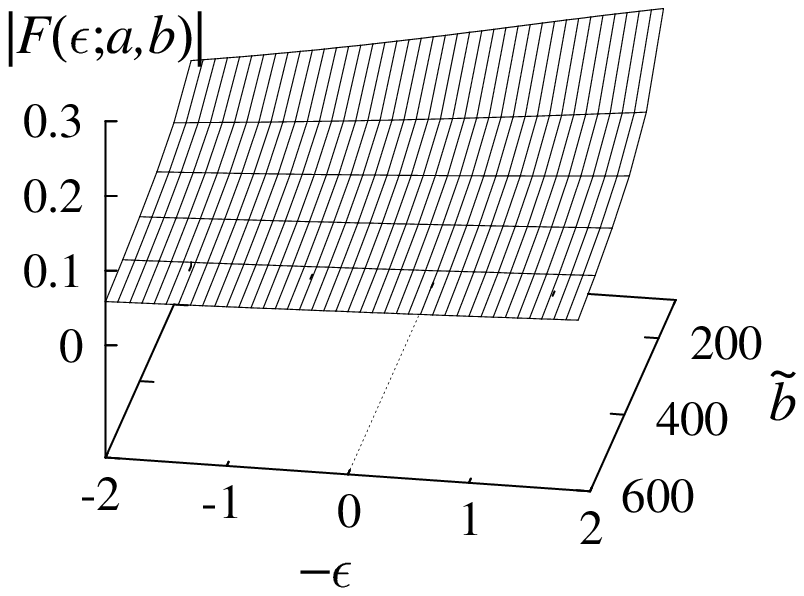} \\
\caption{\label{fig:uaamp6}
Same as Fig.~\ref{fig:uaamp3}, but with the normal form
parameters (\ref{eq:nparam6}) of the hexagonal orbit.}
\end{figure}

The hexagonal orbit encounters also a codimension-two bifurcation of 
the same type as the triangular orbit, as seen in Fig.~\ref{fig:trm6}.
Determining the normal-form parameters in the same manner as above,
we find the values
\begin{equation}
a=\frac{14.0046}{\sqrt{\hbar kR_0}}\,, \quad
b=\frac{414.669}{\hbar kR_0}\,, 
\label{eq:nparam6}
\end{equation}
which are much larger than those for the triangular orbit given in
(\ref{eq:nparam3}). Since they contribute inversely to the local
uniform approximation (\ref{locunif}), this bifurcation has a much
less dramatic influence on the level density than that of the
triangular orbit. In order to demonstrate this more explicitly, we
plot in Figs.~\ref{fig:uaamp3} and \ref{fig:uaamp6} the modulus of 
the integral
\begin{equation}
F(\epsilon;a,b)=\int_0^\infty {\rm d}I J_0(aI^{3/2})\,
e^{i(-\epsilon I-bI^2)} \label{eq:uaamp}
\end{equation}
appearing in (\ref{locunif}). Fig.~\ref{fig:uaamp3} is for the
parameters (\ref{eq:nparam3}) of the triangular and
Fig.~\ref{fig:uaamp6} for the parameters (\ref{eq:nparam6}) of the
hexagonal orbits. We see that for larger $a$ and $b$, this integral
has smaller values and is a more monotonous function of
$\epsilon$. For small $a$ and $b$, however, it takes large values and
exhibits a considerable peak near the bifurcation point $\epsilon=0$.
(Note that the maximum actually occurs slightly {\it after} the
bifurcation, i.e., for $-\epsilon > 0$, as discussed at the end of
Sect.~\ref{sec:orbits}.)  This can be understood as follows. If the
normal-form parameters $a$ and $b$ are small, the action $S$ does not
depend much on $I$ and $\varphi$, and the quasiperiodic orbits around
the central orbit give a more coherent contribution to the integral,
yielding a better restoration of local dynamical symmetry around the
central periodic orbit. Actually, the limit $a\to 0$ is sufficient to
locally restore integrability: the normal form (\ref{eq:normalform})
then only depends on $I$ and for $b\neq0$ describes the bifurcation
of a torus from an isolated orbit (cf.\ \cite{Kaidel1}).

This local dynamical symmetry is equivalent to 
a resonance condition for quasi-tori winding around
the bifurcating periodic orbit over a rather wide region. The
quantization of these quasi-tori yields a number of quasi-degenerate 
levels, seen as the bunches in the spectrum of Fig.\ 
\ref{fig:sps}.  This increases the probability for
small level spacings and renders the NNS distribution more Poisson like.
The considerable changes in the NNS distributions shown in Figs.\ 
\ref{fig:nnsd} and \ref{fig:brody} around $\alpha=0.5$ therefore
account for the emergence of these quasi-resonant tori.

\begin{figure}[t]
\includegraphics[width=\columnwidth]{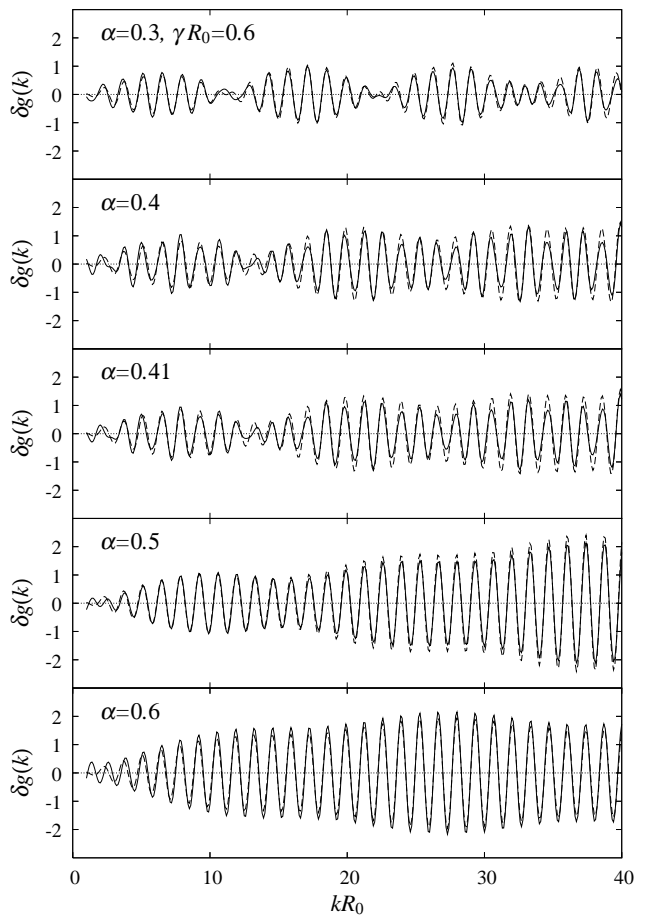}
\caption{\label{fig:density}
Oscillating part of the coarse-grained level density. Solid and dashed
lines represent the semiclassical and the quantum-mechanical results,
respectively.}
\end{figure}

In Fig.~\ref{fig:density} we compare the oscillating part of the 
semiclassical level density with that of the quantum-mechanical
result. In the semiclassical calculation we have included the
primitive diametric, 
triangular, tetragonal, pentagonal and hexagonal orbits as explained
above. We see that, besides the rapid oscillations due to the average 
length of the shortest orbits included (i.e., mainly of the short
diameter orbit), there is a beating pattern
that comes from the interferences between the different orbits.
We note that both amplitudes and phases are nicely
reproduced in the semiclassical result for all deformation parameters
including the bifurcation points $\alpha\simeq0.4$.
For $\alpha=0.5$ and $0.6$, the interference effects are reduced 
and the oscillating pattern is quite regular, indicating the 
dominance of the bifurcating periodic orbits and the accompanying 
quasi-tori which have approximately the same lengths. This
is also seen in the shell-correction energy for $\alpha=0.5$ in 
Fig.\ \ref{fig:sce}, where the spacing $\Delta N$ of the regular 
oscillation corresponds to the length of the dominating triangular
orbit.

\section{Discussion of discrete symmetries}

We conclude by some remarks on the role of discrete symmetries.  In
the above model, the triangular periodic orbit undergoes a bifurcation
in the course of which a pair of three-fold degenerate new orbits are
created. (We only mention the relative degeneracies here; all orbits
have two extra degeneracy factors of 2 due to reflections at the $x$
axis and to time-reversal.)  Stated more generally: the bifurcation
brings about new periodic orbits of reduced symmetry, which have
several degenerate replicas, connected with the symmetry operations
$\mathcal{R}$ and $\mathcal{P}$ in (\ref{eq:dsym}), that give coherent
contributions to the level density. This is one of the reason why we
obtain a strong shell effect due to this bifurcation. If the $C_3$
symmetry is slightly broken, the equilateral triangular orbit will
split into 3 non-equilateral triangular orbits with different lengths
and different stabilities. They will give destructive contributions to
the level density in most of the parameter region.  We would therefore
expect that in billiards with $C_{nv}$ symmetry, the regular polygonal
orbits with $n$ reflections will play the most prominent role, such as
the triangular orbit in the present system.

This behavior can also be predicted from a perturbative trace formula,
developed by Creagh \cite{Crpert}, which describes semiclassically the
breaking of continuous symmetries. In systems with continuous symmetries, 
the leading periodic orbits occur in degenerate families corresponding 
to rational tori. When a perturbation breaks the continuous symmetries, 
the tori are broken into isolated orbits. For weak perturbations, only 
those orbit families (tori) $\xi$ are broken for which the action change 
$\Delta S_\xi$ in lowest order of the classical perturbation 
theory is nonzero. Furthermore, if the perturbed system still has a 
discrete point symmetry, the orbit families which have the same point 
symmetry are in resonance with the perturbation and typically suffer 
the largest first-order action change.
In our billiard system (\ref{eq:shape}), we can treat the deviation 
from the circular billiard, for sufficiently small values of $\alpha$, 
as a perturbation that breaks the continuous U(1) symmetry. For small 
$\alpha$, the shape of the boundary is given by
\begin{equation}
R(\theta)\approx R_0(1-\varepsilon\cos 3\theta)\,,
\end{equation}
and $\varepsilon=\sqrt{3\alpha}/9$ is the appropriate perturbation 
parameter. As we show in Appendix~\ref{app:perturbation}, the length 
of the periodic orbit family $(v,w)$ is changed in first order
of $\varepsilon$ only for $v=3w$, i.e., only triangular orbits are
affected in first-order perturbation theory. The situation is similar 
in a spherical cavity perturbed by deformations with $2^j$-pole 
deformations. For this system it was shown explicitly \cite{Meier} that 
the orbit families with a $j$-fold symmetry obtain 
the largest first-order action changes.  As a consequence of
this symmetry-breaking argument, we understand that the orbits with
the highest point symmetry (namely that of the perturbed system
itself) are those whose stability traces $\Tr M_\xi$ deviate fastest
from their values $\Tr M_\xi=+2$ at $\alpha=0$. Their stable branches
therefore are also the fastest to undergo a bifurcation.

\begin{figure}[tb]
\includegraphics[width=0.8\columnwidth]{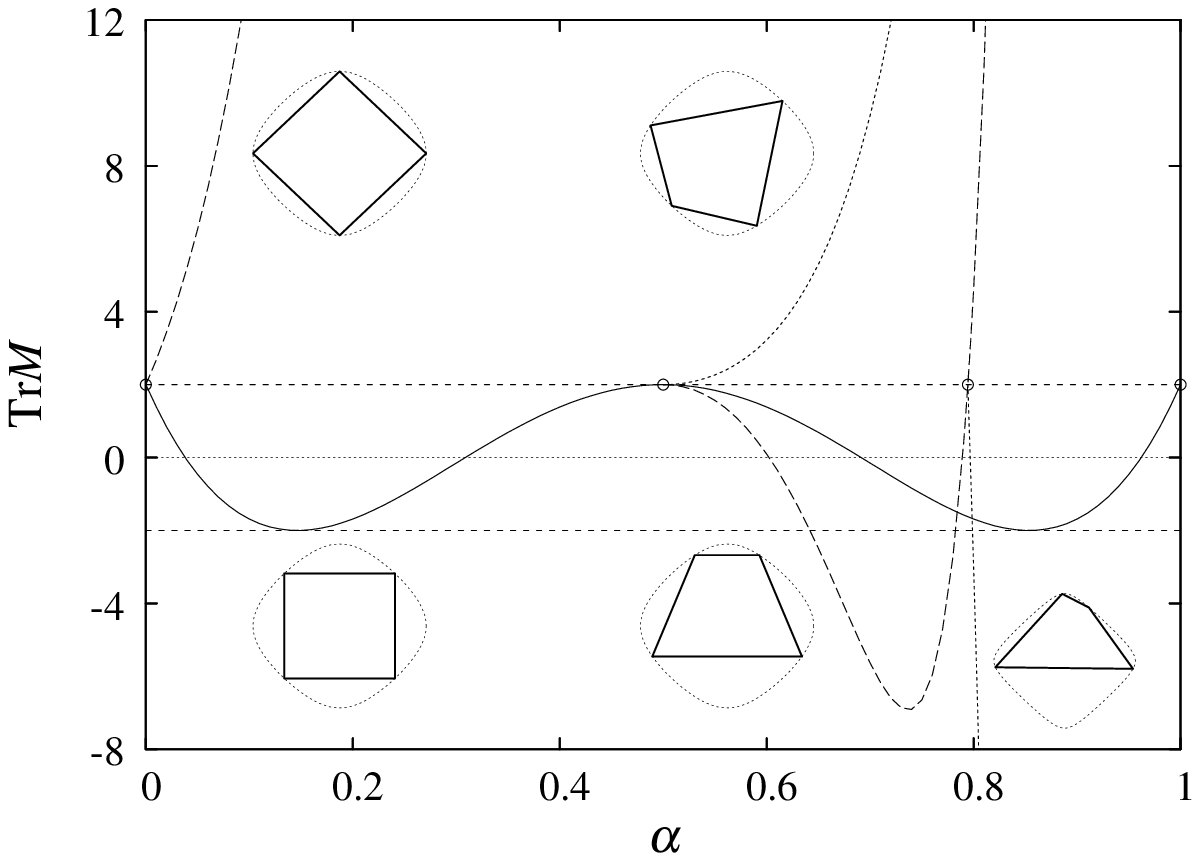} \\
\includegraphics[width=0.8\columnwidth]{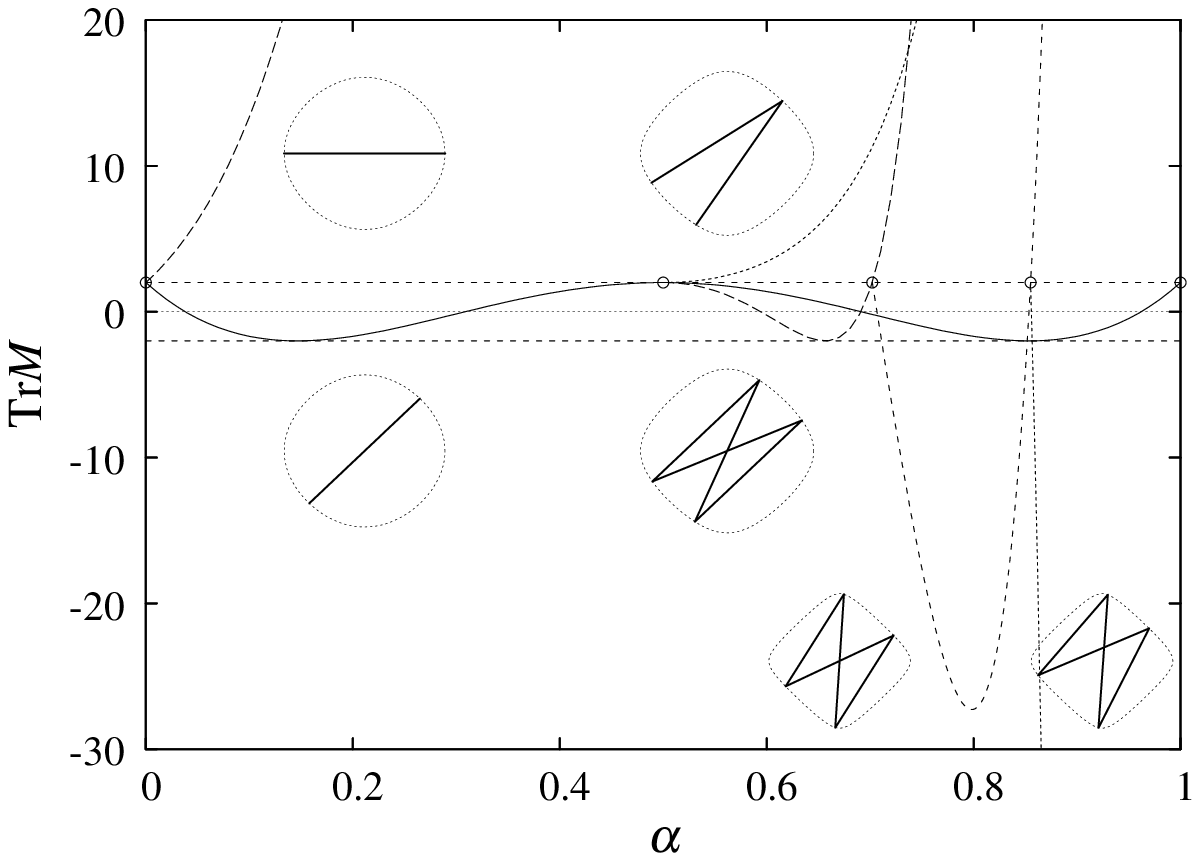}
\caption{\label{fig:trm4}
Trace of stability matrix for the square orbits (upper panel) and
the second repetition of the diametric orbits (lower panel) in the
circular-to-square billiard (\ref{eq:c4vbil}) with $C_{4v}$ symmetry.}
\end{figure}

We finally note that, quite generally, orbits with point symmetries keep 
undergoing bifurcations whereby the new-born orbits at each successive
bifurcation lose one symmetry, until the last stable branch has lost 
all symmetries. (We have no formal proof for this statement, but have 
observed this scenario in many Hamiltonian systems.)  Hereby the stable 
branches survive up to large deformations, giving large contributions 
to the level density, while the unstable branches rapidly turn chaotic 
and their contribution decreases exponentially. In conclusion, orbits 
with high point symmetries tend to give large contributions to the 
level density, even in well-deformed systems, as long as they possess 
discrete symmetries.

\begin{figure}[tb]
\includegraphics[width=\columnwidth]{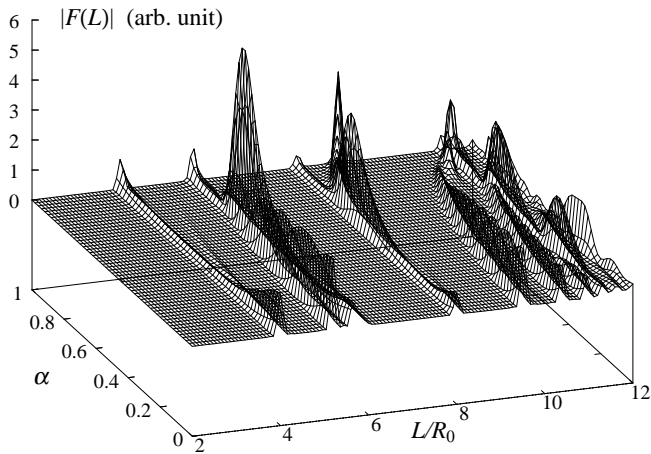}
\caption{\label{fig:ftl4}
Same as Fig.~\ref{fig:ftl3d} for the circular-to-square billiard
(\ref{eq:c4vbil}) with $C_{4v}$ symmetry.}
\end{figure}

To illustrate some of the above statements, let us consider a billiard
system with a different discrete symmetry. The square billiard is 
integrable, and we can smoothly connect square and circular billiards 
by parameterizing the boundary shape $R(\theta)$ as
\begin{equation}
R^2-\frac{\alpha}{8}\frac{R^4(1+\cos 4\theta)-R_0^4}{R_0^2}=R_0^2\,.
\label{eq:c4vbil}
\end{equation}
For small $\alpha$, $R(\theta)$ is expressed by $R(\theta)\simeq 
R_0(1+\varepsilon\cos4\theta)$ with $\varepsilon=\alpha/16$. This 
system obviously has $C_{4v}$ symmetry. $\alpha=0$ corresponds
to the circle with radius $R_0$, and $\alpha=1$ corresponds to 
the square with side length $\sqrt{7/2}R_0$. The system is again 
integrable at these two limits, while it is non-integrable in between.

The square orbit here also has $C_{4v}$ symmetry, and it suffers a
non-generic island-chain bifurcation at $\alpha=0.5$, as seen in the
upper panel of Fig.~\ref{fig:trm4} (see
Appendix~\ref{app:tracem} for the analytical form of its stability
matrix).  The pair of
new-born stable and unstable orbits have lost the $C_{4v}$ symmetry
but still have a reflection symmetry at one axis, they are therefore
doubly degenerate.  The stable branch suffers a pitchfork bifurcation
at $\alpha\sim 0.8$, where it loses the last reflection symmetry. The
diameter orbit (seen in the lower panel of Fig.~\ref{fig:trm4}) has
$C_{2v}$ symmetry; its second repetition suffers an island-chain
bifurcation at $\alpha=0.5$, where a new pair of stable and unstable
branches are born.  At this point, the stable branch
loses its time-reversal symmetry and keeps the $C_{2v}$ symmetry.  The
orbit born at a pitchfork bifurcation near $\alpha\sim 0.7$ loses the
reflection symmetry but still keeps $C_2$ symmetry.  It suffers a
further pitchfork bifurcation near $\alpha\sim 0.85$ and the orbit
born there has lost all symmetries.  The octagonal orbits also show
interesting bifurcation sequences in this system, but their
contributions to the level density are small compared to those of the
above ones. Other short orbits without symmetries are already strongly
chaotic at much smaller values of $\alpha$.

Figure~\ref{fig:ftl4} shows the modulus of the Fourier transform
(\ref{eq:fourier_qm}) of the quantum-mechanical level density of the
circular-to-square billiard (\ref{eq:c4vbil}).  The above bifurcations 
are seen to play a significant role, although their effects are not 
so drastic compared as for the circle-to-triangle billiard, since 
the square and diameter orbits give destructive contributions.

\section{Summary and Conclusions}

We have used periodic orbit theory to analyze a strong enhancement 
of shell effects in a non-integrable billiard system, which is
closely connected to the bifurcation of a periodic orbit with high
point symmetry. The semiclassical trace formula, including the
bifurcating orbits in a global uniform approximation and the other 
isolated orbits with their standard semiclassical Gutzwiller 
amplitudes, describes very well the coarse-grained quantum level 
density of the system. We have shown that the bifurcations play a 
significant role especially when their normal-form parameters are 
small. Under such a condition the quasi-periodic orbits around the 
bifurcating orbit give coherent contributions to the trace integral,
which can be considered as an approximate local symmetry restoration 
around the periodic orbit.  It is accompanied by
a relatively large regular region in the phase space which persists
around $\alpha\simeq0.5$ (cf.\ Fig.\ \ref{fig:pmap}) and is reflected 
by the level statistics (Figs.\ \ref{fig:nnsd}, \ref{fig:brody}), 
and by large shell effects in the spectrum (Fig.\ \ref{fig:sps}) and 
the shell-correction energy (Fig.\ \ref{fig:sce}).

The role of the discrete symmetry is also important for creating a 
large shell effect. The bifurcation of a highly symmetric orbit causes
degenerate symmetry-reduced orbits which give additional contributions 
to the level density.  Through successive bifurcations,
at each of which one symmetry is lost by the new-born orbits, their
stable branches may survive up to rather large deformations until
all symmetries are lost and all remaining branches become
unstable with exponentially decreasing contributions to the level
density. This mechanism is operative in systems with arbitrary point 
symmetries (such as $C_{nv}$ discussed here), in which periodic
orbits of the same symmetries exist.

\acknowledgments
K.A.\ acknowledges financial support 
by the Universit\"atsstiftung Hans Vielberth
and by the Deutsche Forschunsgemeinschaft 
(graduate college 638 ``Nonlinearity and nonequilibrium in condensed
matter'').

\appendix
\section{Analytic expressions for traces of stability matrices}
\label{app:tracem}

If the geometry of a periodic orbit in a billiard is analytically
given, analytic expressions for its stability matrix are often 
available as well. In billiard problems, the stability matrix is 
constructed by alternating products of translation and reflection 
matrices (see, e.g., Appendix B of \cite{CrRoLi}). For a 
periodic orbit with $v$ vertices, it becomes
\begin{equation}
M=M_{\rm ref}(v)M_{\rm tra}(v,v-1)\cdots M_{\rm ref}(1)M_{\rm tra}(1,0)\,,
\end{equation}
where $M_{\rm ref}(i)$ stands for the matrix for reflection at the 
$i$-th vertex, and $M_{\rm tra}(i+1,i)$ that for translation from
the $i$-th to the $i+1$-th vertex, respectively.
They are given analytically by
\begin{gather}
M_{\rm tra}(i+1,i)=\begin{pmatrix} 1 & L_{i+1,i}/p \\
  0 & 1 \end{pmatrix}, \label{eq:m_trans}\\
M_{\rm ref}(i)=\begin{pmatrix} -1 & 0 \\
 2p/\rho_i\cos\varphi_i & -1 \end{pmatrix}, \label{eq:m_refl}
\end{gather}
where $L_{i+1,i}$ is the distance between the vertices $i$ and
$i$+1, $p$ the momentum, $\rho_i$ the curvature radius of the 
boundary at $i$-th vertex, and $\varphi_i$ the reflection angle 
measured from the normal to the boundary at the $i$-th vertex.
For a boundary $R(\theta)$, the curvature radius is given by
\begin{equation}
\rho=\frac{\left(R^2+{R'}^2\right)^{3/2}}{R^2+2{R'}^2-RR''}
  \label{eq:rho_formula},
\end{equation}
where primes indicate derivatives with respect to $\theta$.

We first apply this to the equilateral triangular orbit 3A in the
billiard system (\ref{eq:shape}). Its vertices are located at $\theta=0, \pm
2\pi/3$, and the curvature radius at these points is
\begin{equation}
\rho=\frac{(3-\gamma^2)\gamma}{2(4\gamma^2-3)}R_0, \label{eq:rho_triangle}
\end{equation}
where $\gamma$ is the distance between the origin and the reflection
points in units of $R_0$; it is a function of the deformation parameter
$\alpha$ given implicitly by the equation
\begin{equation}
\gamma^2+\frac{2\sqrt{3\alpha}}{9}\gamma^3=1\,. \label{eq:gamma_alpha}
\end{equation}
Since the reflection and translation matrices for all vertices and sides
are equal, the stability matrix is simply given by
$M=(M_{\rm ref}M_{\rm tr})^3\equiv M_1^3$.
Using the symplectic nature of the matrices $M$, we can express the trace of
the full $M$ in terms of the trace of $M_1$:
\begin{equation}
\Tr M=\Tr (M_1)^3=(\Tr M_1)^3-3\Tr M_1\,.
\end{equation}
Inserting $L_{i+1,i}=\sqrt{3}\gamma R_0$, $\varphi_i=\pi/6$ and
(\ref{eq:rho_triangle}) into (\ref{eq:m_trans}) and (\ref{eq:m_refl}), 
we get $\Tr M_1=(34\gamma^2-30)/(3-\gamma^2)$.
At the bifurcation point, $\Tr M_1=-1$ and thus
$\gamma=3/\sqrt{11}$.  Inserting this into (\ref{eq:gamma_alpha}),
we get the deformation
\begin{equation}
\alpha_{\rm 3A}=11/27=0.407407...
\end{equation}

Bifurcations of the square orbit 4A and the linear orbit 2A in the
circle-to-square billiard system (\ref{eq:c4vbil}) are analyzed
in the same way.  For both orbits, we have
$\Tr M_1=2-4\alpha$.  Thus, orbit 4A and the 2nd repetition of
orbit 2A have the same value of $\Tr M$ at all deformations.
The bifurcation deformation becomes, from $\Tr M_1=0$,
\begin{equation}
\alpha_{\rm 4A}=\alpha_{{\rm 2A}^2}=1/2\,.
\end{equation}

\section{Classical perturbation of the circular billiard}
\label{app:perturbation}

In this appendix we calculate the change of the periodic orbit lengths 
in the circular billiard, when it is perturbed by a deformation
\begin{equation}
R(\theta)=R_0(1-\varepsilon\cos n\theta)\,.
\label{eq:c1}
\end{equation}
We consider the primitive periodic orbit family $(v,w)$, with mutually
prime integers $v$ and $w$ representing the number of vertices and the
winding number around the origin, respectively.  Due to the
deformation, the positions of the vertices $\theta_i$ $(i=1,\cdots,v)$
are shifted by $\delta\theta_i$, which are of order $\varepsilon$.  The
distance between successive vertices is given by
\begin{gather}
L_{i+1,i}^2=\;R_i^2+R_{i+1}^2
 -2R_iR_{i+1}\cos(\theta_{i+1}-\theta_i)\,,\nonumber \\
R_i=R(\theta_i), \quad \theta_i=\theta_0+i\frac{2w\pi}{v}+\delta\theta_i\,,
\end{gather}
where one of the two branches has $\theta_0=0$ and the other branch
has $\theta_0=\pi/l_{nv}$, $l_{nv}$ being L.C.M. of $n$ and $v$.
Inserting (\ref{eq:c1}) and keeping only terms up to
first order in $\varepsilon$, we obtain
\begin{align}
& \Delta L_{i+1,i}\approx R_0\Biggl\{
\cos\frac{w\pi}{v}
 (\delta\theta_{i+1}-\delta\theta_i)~~~~ \nonumber\\
&\qquad\quad\quad\; -\varepsilon\sin\frac{w\pi}{v}
 \biggl[\cos\left(n\theta_0+\frac{2niw\pi}{v}\right) \nonumber\\
&\hspace{22mm}+\cos\left(n\theta_0+\frac{2n(i+1)w\pi}{v}\right)
 \biggr]\Biggr\}.
\end{align}
Therefore, the changes of the periodic orbit lengths are
\begin{align}
\Delta L_{vw}&=\sum_{i=1}^v \Delta L_{i+1,i} \nonumber\\
&\approx -2\varepsilon R_0\sin\frac{w\pi}{v}\sum_{i=1}^v
 \cos\left(n\theta_0+\frac{2niw\pi}{v}\right).
\end{align}
The sum on the right-hand side has a non-vanishing value only for
$n/v=t$, $t$ being an integer; then it becomes
\begin{align}
\Delta L_{vw}=-2v\varepsilon R_0\sin\frac{w\pi}{v}\cos(n\theta_0)
 =\mp 2v\varepsilon R_0\sin\frac{w\pi}{v}\,.
\end{align}
All other periodic orbit families suffer a change of length only 
at higher orders of $\varepsilon$.


\begin{thebibliography}{99}
\bibitem{note} The generic touch-and-go bifurcation is period
               tripling \cite{Ozorio}, and the factor $n_r$ in 
               (\ref{eq:phtrace}), which was derived for generic 
               bifurcations, is the repetition number $n_r=3$. Due to the
               $C_{3v}$ symmetry of the present system, the bifurcation
               is non-generic and isochronous. The factor $n_r$ here 
               is the discrete symmetry degeneracy $f=3$ of the 
               new orbits, since it generically counts the number of 
               fixed points of each satellite orbit involved 
               in the bifurcation.
\end{thebibliography}
\end{document}